\documentclass[12pt]{article}
\input{psfig}

\begin{document}
\begin{titlepage}
\begin{center}
{\Large {\bf
 STRUCTURE OF THE ENERGY LANDSCAPE OF SHORT PEPTIDES
}}
\end{center}
\vskip 1.0cm
\centerline{ HANDAN ARKIN and TARIK \c{C}EL\.{I}K$^\dagger$}
\vspace*{-0.2cm}
\centerline{\it Hacettepe University, Physics Department,
          Ankara, Turkey}
\centerline{$^\dagger$ {\it E-mail: tcelik@hacettepe.edu.tr}}
\vskip 0.3cm

\begin{abstract}
 We have simulated, as a showcase, the pentapeptide Met-enkephalin
(Tyr-Gly-Gly-Phe-Met) to visualize the energy landscape and investigate the
conformational coverage  by the multicanonical method.
We have obtained a three-dimensional topographic picture of the
whole energy landscape 
 by plotting the histogram with respect to
energy(temperature) and the order parameter, which gives the
degree of resemblance of any created conformation with the
global energy minimum (GEM).

\end{abstract}

\end{titlepage}

\section{Introduction}

Biological macromolecules such as  proteins
have a well defined 3D structure which  is essential for 
their biological activity. Therefore, predicting the protein's structure by 
theoretical/computational
methods is an important goal in structural biology.~\cite{CHEMREV} 

\bigskip 

The configuration space of peptide's and protein's presents a complex energy
profile consisting of  
tremendous number of local minima;  their  basins of attraction
were  called localized microstates. The energy profile also 
contains  larger
potential energy wells defined over wide microstates
(e.g., the protein's fluctuations around its
averaged structure),
each including many localized ones.~\cite{methodology}

\bigskip

Because of energy barriers, the commonly used
thermodynamic simulation techniques, such as
the Metropolis Monte Carlo (MC)~\cite{MC} 
and molecular dynamics (MD) ~\cite{MD} are not very efficient 
in sampling a rugged landscape. Thus, the  molecule remains in its 
starting wide microstate or move to a 
neighbor wide microstate, but in practice will hardly reach the most 
stable one. The system may occur to be trapped in a basin for a long time,
which results in non-ergodic behavior.  Therefore, 
developing  simulation methods that 
lead to an efficient crossing of the energy barriers
has been a long standing challenge.

\bigskip

The topography of the energy landscape,
especially near the global minimum is of particular importance, because
the potential energy surface defines the behavior of the system. 
Methods for searching energy surfaces are proposed~\cite{Ch01},
energy landscape perspectives are investigated ~\cite{OnLu97} and the
fractal dimensions are studied~\cite{Th99}.
The essence of a funnel structure of energy landscape at
some fixed temperatures has recently been shown by Hansmann and 
Onuchic~\cite{HaOn99}.
Consequently, a visualization of the whole rugged landscape covering the
entire energy and temperature ranges would be helpful to develop
methods allowing one to survey the distribution of structures in conformational spaces.  
Such a goal can be achieved within the multicanonical ensemble approach.

\bigskip

An ideal simulation scheme should freely visit the entire configuration space and
predominantly sample the significant conformations. 
The trapping problem of the MC and MD methods can 
be alleviated to a large extent, by the multicanonical MC method (MUCA)  
~\cite{BeNe91,BeCe92,Be99}, which  was applied initially 
to lattice spin models and its relevance for complex systems was first 
noticed in 
Ref. \cite{BeCe92}.  Application of the multicanonical approach
 to peptides was pioneered by 
Hansmann and Okamoto~\cite{HaOk93} and followed by 
others ~\cite{HaSc94}; simulations of protein folding with
MUCA and related generalized ensemble methods are reviewed in 
Refs.~\cite{HaOk99rev} and ~\cite{Ok00rev}.

\bigskip

\section{The Model}

The multicanonical ensemble  based on a probability 
function in which 
the different energies are equally probable.
However, implementation of  MUCA is not straightforward
because the density of states  $n(E)$ is unknown {\it a priori}.
In practice, one only needs  to know the weights  $\omega$, 
\begin{equation}
 w(E) \sim 1/n(E) = \exp [(E-F_{T(E)})/k_BT(E)].
\end{equation}

These weights are calculated in the 
first stage of simulation  process  by an iterative  procedure 
in which
the temperatures $T(E)$ are built recursively together with 
the microcanonical free energies $F_{T(E)}/k_BT(E)$,  up to an 
additive constant. 
The iterative  procedure is
followed by a long production run based on the fixed $w$'s where 
equilibrium configurations are sampled. Re-weighting techniques 
(see Ferrenberg and Swendsen~\cite{FeSw88} and literature given in 
their second reference)
enable one to obtain Boltzmann averages of various 
thermodynamic properties over a large range of temperatures.

\bigskip

As pointed out above, calculation of the {\it a priori} unknown MUCA weights   
is not trivial, requiring an experienced human intervention.  
For lattice models, this problem was addressed in a
sketchy way by Berg and \c{C}elik~\cite{BeCe92} and later by
Berg~\cite{Be98}.
An alternative way is to establish an automatic process by
incorporating the statistical errors within the recursion
procedure.  
The automatic procedure was tested successfully ~\cite{Ya00} as applied to  
models of the pentapeptide Leu-enkephalin 
(H-Tyr-Gly-Gly-Phe-Leu-OH) described by the ECEPP/2 potential energy 
function~\cite{ECEPP}.

\bigskip 

In this work, as in our previous one,~\cite{Ya00} Met-enkephalin is 
modeled  by the ECEPP/2 potential, which assumes a rigid geometry,  
and is based on non-bonded, Lennard-Jones, 
torsional, hydrogen-bond, and electrostatic potential terms  
with the dielectric constant $\epsilon=2$. This potential energy is implemented
into the software package SMMP ~\cite{SMMP}. We further fix peptide bond angles
$\omega$ to their common value $180^{\rm o}$, which leaves us with 19 dihedral angles
as independent degrees of freedom ( $n_F = 19$ ). We have also simulated  Met-enkephalin
with variable  peptide bond angles, for which the distribution of conformations are
included in Table I. 

\bigskip

\section{Results and Discussions}
 
We first carried out canonical (i.e., 
constant $T$) MC simulations at the relatively high temperatures and
MUCA test runs which enabled us to determine the required energy
ranges. Then we preformed full simulation which cover the
high temperature region up to $T_{\max}=1000\,$K reliably. 
The energy range was divided into 
31 bins of $1\,$kcal/mol each, covering the range $[20,-11]\,$kcal/mol. 
The lowest energy encountered was $-10.75\,$kcal/mol and 
$T_{\max}=1000\,$K was also used above $20\,$kcal/mol.  
At each update step, a 
trial conformation was obtained by changing {\it one} dihedral angle 
at random within  the range [$-180^{\rm o};180^{\rm o}]$, followed by 
the Metropolis test and an update of the suitable histogram. The 
dihedral angles were always visited in a predefined (sequential) 
order, going from Tyr to Met; a cycle of $N$ MC steps ($N$=19) is 
called a sweep. 
The weights  were built after $ m = 100 $ recursions during a 
long {\it single} simulation, where the parameters $b_i$ and $a_i$ 
were iterated every 5000~sweeps. 

\bigskip

For peptides it is not only of 
interest to obtain thermodynamic averages and fluctuations at different  
temperatures but also to find the most stable regions in conformational
space populated by the molecule. In the organic chemistry community
conformational search methods have been developed and attempts have
been made to find the global energy minimum (GEM) and all the energy minimized 
conformations in certain energy ranges above the GEM (see Ref.2(b), and 
references cited therein). 

\bigskip

The lowest energy conformation (our suspected GEM) was found at
$ E = - 10.75\,{\rm kcal/mol}$. 

Here we define, following Hansmann et. al. ~\cite{HaOn99}, an order parameter (OP) 
\begin{equation}
OP = 1 -\frac{1}{90~n_F} \sum_{i=1}^{n_F} |\alpha_i^{(t)}- \alpha_i^{(RS)}|~,
\end{equation}
where $\alpha_i^{(RS)}$  ve $\alpha_i^{(t)}$
are the dihedral angles of the reference state (which is taken as GEM) and
of the considered configuration, respectively. The difference  
$\alpha_i^{(t)}- \alpha_i^{(RS)}$  is always in the interval 
$[-180^{\circ},180^{\circ}]$, which in turn gives for peptides 
\begin{equation}
0 \le ~<OP>_T~ \le 1
\end{equation} 

Figure.1 shows the energy landscape obtained by the multicanonical simulation run 
of one million sweep plotted against energy and the order parameter. Here, 
we would like to point out that the utilized data is obtained by sampling
of the conformational space and no minimization procedure is applied. 
At high temperatures, where the peptide is in the random coil state,
the histogram looks as one gaussian-like peak centered around
the value of the order parameter $ OP \sim 0.3 $. When the temperature
is lowered, first a transition from the state of random coil to globular 
structure is expected. In Figure.2 we show the same energy landscape of Fig.1(b)
 by
grouping the conformations of $1$ kcal/mol interval in energy. Curve a) denotes 
the energy interval $-1$ kcal/mol $\le E \le 0$ kcal/mol, which corresponds 
after re-weighting to the temperature interval $315$ K $\le T_a \le 330$ K.
At this temperature, the energy landscape starts deviating from a smooth
surface and develops a shoulder. We identify this 
temperature as the starting of forming a structure rather than a random coil. 
Further down in 
energy (temperature), the newly developing branch 
becomes more populated. At the temperature $215$ K $\le T_b \le 230$ K  
denoted by the curve
b), the energy landscape displays a typical structure bifurcating into
two branches of almost equal height. From there 
on, the branch having larger values of the order parameter wins and more
conformations populate that section of the conformational space.  
Our estimate of $T_a $ and $T_b$ from the topographic structure of 
the potential energy surface of
Met-enkephalin are very close to the values of the collapse temperature
 $T_\theta = 295 \pm 20$ K and the folding temperature
 $T_f = 230 \pm 30$ K, respectively, determined by Hansmann
 et al ~\cite{HaMa97}. 
We observe a third temperature denoted by the curve c) in Fig.2 where the 
glassy behavior sets in and many valley structure of the landscape 
become clearly pronounced. For our simulated peptide sample Met-enkephalin,
this temperature is in the range $155$ K $\le T_c \le 185$ K. Below this
temperature, one observes the appearances of multiple  valleys which are well 
separated by high energy barriers.
The valley at the far-out end of the order parameter scale having the 
conformations with the value of the order parameter in the range 
$ 0.98 \le OP \le 1$ contains the global energy minimum (GEM), respect to
which the order parameter is evaluated. The temperature $T_c$ seems to 
correspond  to the glass transition temperature estimate 
 of $T_g = 180 \pm 30$ K, which value is based on the fractal 
 dimension estimates.~\cite{Th99}
 In Fig.3 we plotted all the 
conformations found with energy $ E \le -10.5$ kcal/mol with respect to the order parameter.
Their number is 3587 conformations in one production run of one million sweeps.
As  clearly seen from Fig.3 that the conformations in this energy range 
are localized in
one of the four valleys, which are identified by the value of 
their order parameter
$OP \sim 0.80, 0.87, 0.92$ and $0.98$. The conformations in the neighborhood
of the GEM take place within the same wide microstate of the GEM but they
are grouped into local microstates, each of which are one of the above mentioned
valleys.  The small differences in values of OP comes from the differences in
side-chain angles. We observe no conformation anywhere outside the definite 
valleys when the energy is less than about 1 kcal/mol above the GEM.

\bigskip

The number of conformations found in energy bins of $1$ kcal/mol, which 
were plotted in Fig.2, appear in Table I. The lowest bin is $0.75$ kcal/mol 
and includes the
GEM. The table displays the distribution of sampled conformations according to 
the order parameter values, namely the distribution with respect to 
how far they are in configuration space from the global energy minimum.
We also included in Table I the same distribution obtained in our simulation
of Met-enkephalin for the case of variable peptide-bond angles $\omega$.  

\smallskip

In conclusion, we have simulated the pentapeptide Met-enkephalin by utilizing
the multicanonical ensemble approach and investigated the structure of the 
rugged energy landscape in the configurational space. We were able to display the 
distribution of 
 at all temperatures from a single simulation 
and estimate the critical
temperatures such as the collapse temperature, the folding temperature and 
the glass transition temperature. Such a visualization would be helpful
in designing algorithms for efficient sampling of conformational space. 

\smallskip
\section{Acknowledgements}

This work has been supported by the
Hacettepe University Research Fund through project number
00.01.602.001. We are gratefull to U.H.E. Hansmann for
providing us the SMMP ~\cite{SMMP}.

\begin{table}
\vspace{0.3cm}
\caption{Number of conformations in energy bins of 1 kcal/mol.}
{\centering \begin{tabular}{|c|cccc|c|}
\hline 
ENERGY&
&
OVERLAP&
&
&
TOTAL\\
\hline 
\hline 
Fix \( \omega  \)&
1.0-0.9&
0.9-0.8&
0.8-0.7&
0.7-0.6&
CONF.\\
\hline 
\hline 
-10.75 to -10.0&
3282&
3935&
3073&
2779&
15327\\
\hline 
-10.0 to -9.0&
1001&
3530&
4925&
4475&
28088\\
\hline 
-9.0 to -8.0&
467&
2332&
4003&
3979&
26220\\
\hline 
-8.0 to -7.0&
190&
1460&
3150&
3488&
24139\\
\hline 
-7.0 to -6.0&
90&
897&
2515&
3290&
22497\\
\hline 
Variable \( \omega  \)&
&
&
&
&
\\
\hline 
-12.21 to -12.0&
23&
25&
-&
-&
48\\
\hline 
-12.0 to -11.0&
6380&
7568&
302&
197&
14457\\
\hline 
-11.0 to -10.0&
7600&
21199&
4775&
2784&
37107\\
\hline 
-10.0 to -9.0&
2700&
9956&
3959&
3456&
28430\\
\hline 
-9.0 to -8.0&
600&
3107&
2390&
3137&
2644\\
\hline 
\end{tabular}\par}
\vspace{0.3cm}
\end{table}

\begin{figure}
\psfig{figure=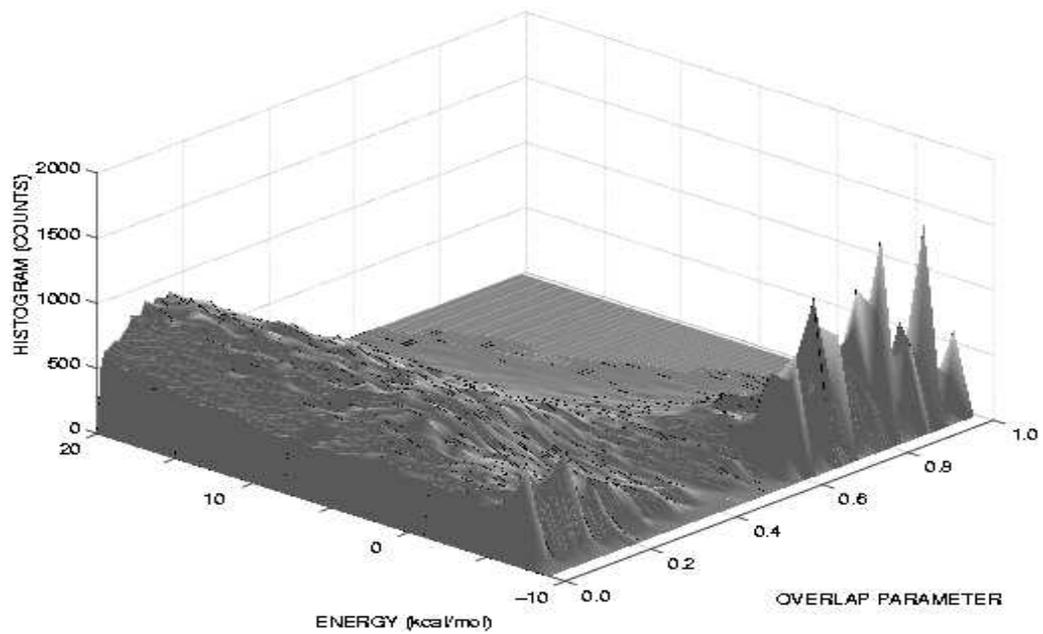,height=9cm,width=14cm}
\vspace{1pc}
\psfig{figure=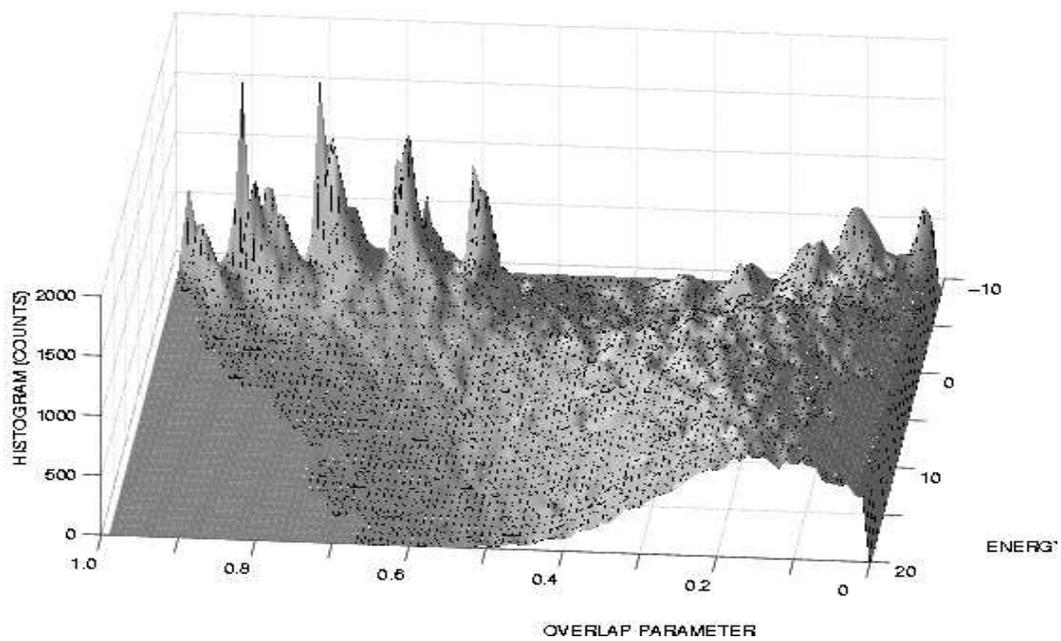,height=9cm,width=14cm}
\caption{Energy surface in configuration space of Met-enkephalin
          viewed from different angles. }
\end{figure}
\begin{figure}
\hspace{0.1cm}
\psfig{figure=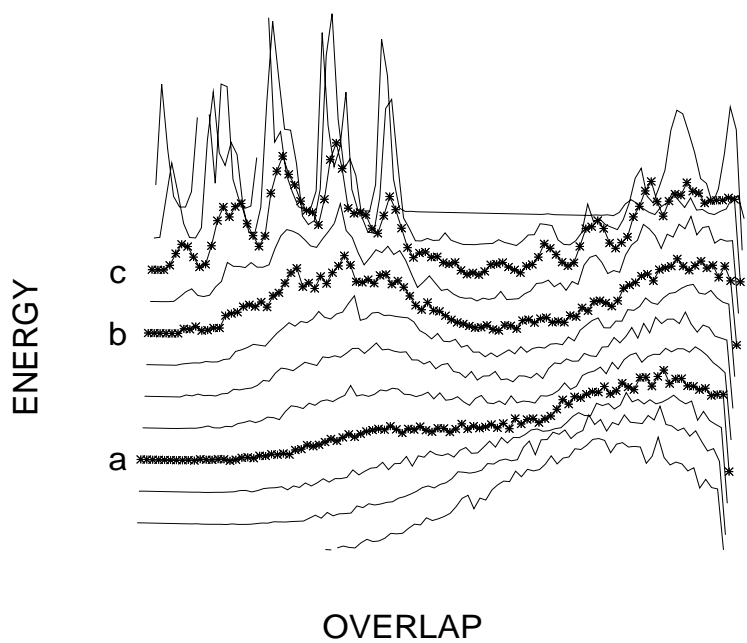,height=10cm,width=15cm}
\caption{Same as Fig.1(b), plotted by
grouping the conformations of $1$ kcal/mol interval in energy.}
\end{figure}
\begin{figure}
\hspace{0.1cm}
\psfig{figure=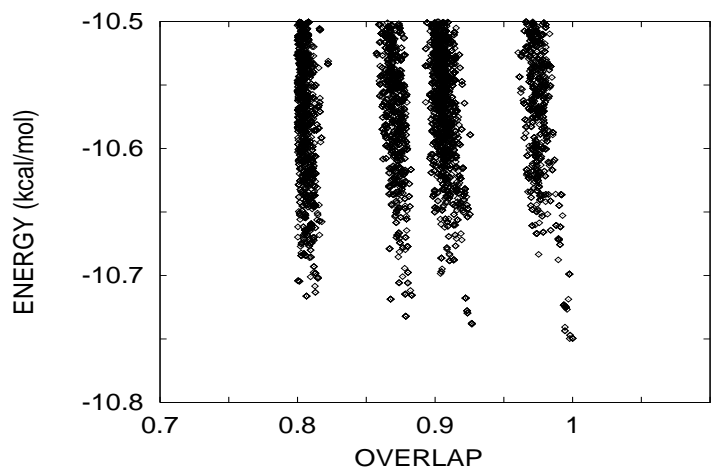,height=6cm,width=10cm}
\caption{Distribution of microstates with $ E \le -10.5 $ kcal/mol with respect to the overlap parameter.}
\end{figure}

\end{document}